\documentclass[apj]{emulateapj}

\begin{document}

\submitted{Accepted for publcation to the Astropysical Journal on June 2, 2014}

\shorttitle{Modeling X-rays from PSR J1852+0040}
\shortauthors{Bogdanov}

\title{Modeling the X-rays from the Central Compact Object PSR J1852+0040 \\
  in Kesteven 79: Evidence for a Strongly Magnetized Neutron Star}

\author{Slavko Bogdanov} 

\affil{Columbia Astrophysics Laboratory, Columbia University, 550 West 120th Street, New York, NY 10027, USA\\ slavko@astro.columbia.edu}

\begin{abstract}
I present modeling of the X-ray pulsations from the central compact
object (CCO) PSR J1852+0040 in the Galactic supernova remnant Kesteven
79. In the context of thermal surface radiation from a rotating
neutron star, a conventional polar cap model can reproduce the broad,
large-amplitude X-ray pulse only with a ``pencil plus fan'' beam
emission pattern, which is characteristic of strongly magnetized
($\gtrsim$$10^{12}$ Gauss) neutron star atmospheres, substantially
stronger than the $\sim$$10^{10}$ Gauss external dipole field inferred
from the pulsar spin-down rate. This discrepancy can be explained by
an axially displaced dipole.  For other beaming patterns, it is
necessary to invoke high-aspect-ratio emitting regions that are
greatly longitudinally elongated, possibly due to an extremely offset
dipole.  For all assumed emission models, the existence of strong
internal magnetic fields ($\gtrsim$$10^{14}$ Gauss) that
preferentially channel internal heat to only a portion of the exterior
is required to account for the implied high-temperature contrast
across the stellar surface.  This lends further observational evidence
in support of the ``hidden'' strong magnetic field scenario, in which
CCOs possess strong submerged magnetic fields that are substantially
stronger than the external dipole field, presumably due to burial by
fallback of supernova ejecta.  I also conduct phase-resolved X-ray
spectroscopy and find no evidence for prominent spin-phase-dependent
absorption features that could be produced by cyclotron
absorption/scattering.
\end{abstract}


\section{Introduction}
Central compact objects (CCOs) constitute a group of X-ray-emitting,
radio-quiet neutron stars found near the centers of supernova remnants
(SNRs).  To date, X-ray pulsations have been firmly detected from only
three CCOs. They have relatively long spin periods ($0.1-0.4$ s), and
long-term monitoring shows that their period derivatives
($\dot{P}\equiv {\rm d}P/{\rm d}t$) are remarkably small, suggesting
weak surface magnetic fields\footnote{By assuming dipole spin-down,
  the surface field strength can be inferred from
  $B_{\rm surf} \equiv3.2\times10^{19}(P\dot{P})^{1/2}$\,G, where $P$ is in
  seconds.}. See \citet{Halpern10}, \citet{Got09,Got13}, and
\citet{Ho13} for observations and overview of related theory.  Due to
the limited sample of CCOs that exhibit X-ray pulsations, the physical
mechanism responsible for their X-ray emission is not well understood,
and their active lifetime and long-term evolution are poorly
constrained.  It remains unclear if CCOs are active radio pulsars
beamed away from us or if the radio emission mechanism is
intrinsically inoperative.  Since CCOs are associated with very young
SNRs, their nature and evolution are highly relevant to the neutron
star production rate and the physics underlying the diversity of
neutron stars produced by core collapse.

The compact X-ray source CXOU J185238.6+004020 was discovered in the
center of the SNR Kesteven 79 by \citet{Seward02}. Subsequently,
\citet{Got05} discovered 105 ms pulsations from this CCO, now named
PSR J1852+0040, establishing it as a young neutron star. A dedicated
long-term X-ray timing campaign of PSR J1852+0040 facilitated the
first definite measurement of the spin-down rate of a CCO pulsar by
\citet{Halpern10}.  The measurements of $P= 0.105$ s and $\dot{P}=8.7
\times10^{-18}$ s s$^{-1}$, imply in the dipole spin-down formalism a
surface magnetic field strength of only $B_{\rm surf} =
3.1\times10^{10}$ G; on this basis, it has been termed an
``anti-magnetar''. With a bolometric luminosity of $3.0\times10^{33}$
erg s$^{-1}$, an order of magnitude higher than its spin-down power
$\dot{E} \propto \dot{P} P^{-3} = 3.0 \times 10^{32}$ erg s$^{-1}$,
the X-ray radiation from PSR J1852+0040 is clearly not powered by the
rotational kinetic energy of the star, thus requiring an additional
energy source, such as residual cooling or low-level accretion.

 The thermal X-ray emission from PSR J1852+0040 is characterized by a
 single unusually broad pulse, with a very high pulsed fraction of
 $64\%\pm2\%$. X-ray observations spanning nearly five years are
 consistent with steady flux. Fitting of the X-ray spectrum to two
 blackbodies finds small emitting radii \citep[$R_1 = 1.9$ km and $R_2
   = 0.45$ km, for components of $kT_1 = 0.30$ keV and $kT_2 = 0.52$
   keV, respectively;][]{Halpern10}.  Such small, hot regions are
 common among CCOs and are at odds with the inferred magnetic field
 strength since highly non-uniform surface temperature is usually
 attributed to the effects of much stronger magnetic fields. Thus it
 is unclear whether CCOs are intrinsically weakly magnetized neutron
 stars or whether they possess substantially stronger internal
 magnetic fields than the measured $\sim$$10^{10}$ G surface dipole
 field.  This fundamental question regarding the nature of CCOs has
 generated a flurry of theoretical efforts aimed at constraining the
 key physics and evolutionary fate of these enigmatic objects
 \citep[see, e.g.,][]{Ho11,Shab12,Vig12,Ber13,Per13}.  It is highly
 likely that the heat distribution on the stellar surface closely
 traces the magnetic field structure.  Therefore, constraining the
 surface emission properties and heat distribution of PSR J1852+0040
 can help resolve this essential mystery of CCOs.

In this paper, I present modeling of the pulsed thermal X-ray emission
from PSR J1852+0040 aimed at constraining key aspects of CCO physics
based on the extensive set of archival \textit{XMM-Newton}
observations.  The work is organized as follows. In \S2 I summarize
the archival data set and the data reduction procedures. In \S3 I
describe the numerical model employed in this study, while in \S4 I
show the results of the modeling. In \S5 I present a
pulse-phase-resolved spectroscopic analysis. I discuss the implications
of the results in \S5 and offer conclusions in \S6.

\begin{deluxetable}{rcc}
\tablewidth{0pt}
\tablecaption{\textit{XMM-Newton} X-ray Timing Observations of PSR J1852+0040.}
\tablehead{
 \colhead{ObsID} & \colhead{Date} & \colhead{Exposure\tablenotemark{a}}  \\
 \colhead{}      & \colhead{(UT)} & \colhead{(ks)}}
\startdata
0204970201   & 2004 Oct 18 & 30.6  \\
0204970301   & 2004 Oct 23 & 30.5  \\
0400390201   & 2006 Oct 08 & 29.7  \\
0400390301   & 2007 Mar 20 & 30.5  \\
0550670201   & 2008 Sep 19 & 21.2  \\
0550670301   & 2008 Sep 21 & 31.0  \\
0550670401   & 2008 Sep 23 & 34.8  \\
0550670501   & 2008 Sep 29 & 33.0  \\
0550670601   & 2008 Oct 10 & 36.0  \\
0550671001   & 2009 Mar 16 & 27.0  \\
0550670901   & 2009 Mar 17 & 26.0  \\
0550671201   & 2009 Mar 23 & 27.3  \\
0550671101   & 2009 Mar 25 & 19.9 \\
0550671301   & 2009 Apr 04 & 26.0 \\
0550671901   & 2009 Apr 10 & 30.5 \\
0550671801   & 2009 Apr 22 & 28.0
\enddata
\tablenotetext{a}{Total observing time not corrected for the 29\% dead time of the EPIC pn small window mode.}									
\label{logtable}							 
\end{deluxetable}

\section{Data Reduction}
I have retrieved the set of 16 archival \textit{XMM-Newton} European
Photon Imaging Camera (EPIC) pn \citep{struder01} observations of PSR
J1852+0040, for a combined 327 kiloseconds of net exposure time (see
Table 1). All exposures were obtained in small window mode, which
affords a 5.7 ms time resolution but at a cost of 29\% dead time
during which no X-ray events are recorded.  Each ODF data set was
reprocessed with the SAS\footnote{The \textit{XMM-Newton} SAS is
  developed and maintained by the Science Operations Centre at the
  European Space Astronomy Centre and the Survey Science Centre at the
  University of Leicester.}
version xmmsas\_20120523\_1702-12.0.0 {\tt epchain} pipeline to ensure
that the latest calibration products and clock corrections (including
leap seconds) are applied. The data were then filtered using the
recommended standard pattern, flag, and pulse invariant values.  None
of the observations exhibit instances of high background flares.

For the purposes of the analyses presented below, the photon arrival
times from each observation were translated to the solar system
barycenter with the SAS {\tt barycen} tool assuming the DE405 solar
system ephemeris. The corrected arrival times were folded coherently
at the pulsar period using the ephemeris presented in
\citet{Halpern10}.  Relative to this previous analysis, which was
based on the same data set but processed with SAS version
xmmsas\_20060628\_1801-7.0.0, there is a systematic offset of $+$2.9 ms
in all photon arrival times. This difference can be attributed to
improvements in the \textit{XMM-Newton} clock corrections. For all
practical purposes this discrepancy is negligible and does not affect
the results and conclusions of \citet{Halpern10}.

The X-ray events from the pulsar was extracted from a 12$''$ radius
circle. This relatively small region was chosen so as to minimize the
contribution of the diffuse emission from the supernova remnant. An
important prerequisite for the pulse profile analysis described below
is a reliable estimate of the background level at the source position.
However, due to the relatively bright diffuse emission, coupled with
the complicated morphology of the portion of the remnant that falls within
the small-window mode \textit{XMM-Newton} images, there is no obvious
choice for a background extraction region. For this purpose, I take
advantage of the sub-arcsecond resolution of the archival 30 ks
\textit{Chandra} ACIS-S image of Kes 79 (ObsID 1982) to identify a
representative background region. I estimated the background level at
the pulsar position by extracting counts from an annulus with inner
radius of $2''$, beyond which the point source emission becomes
negligible, and outer radius of $12''$. The resulting value was used
to identify a larger background region with a matching surface
brightness (i.e., count rate per unit area).

\section{The Numerical Model}

\subsection{System Geometry and General Relativity}
To study the pulsed X-rays from PSR J1852+0040, I employ a numerical
model of surface emission from a neutron star assuming a Schwarzschild
metric to describe the properties of the space-time in the vicinity of
the star. It follows the basic formalism first presented by
\citet{Pech83} and used in a host of subsequent works
\citep[e.g.,][]{Ftac86,Riff88,Mill98,Crop01,Wein01,Belo02,Pou03,Vii04,Got10}.
I represent the thermally-emitting pulsar by a neutron star of mass
$M$, radius $R_{NS}$, spin period $P$, with different arrangements of
X-ray-emitting surface elements on an otherwise cold neutron star. The
surface normal of each element is at a position angle $\alpha$
relative to the spin axis, while the line of sight to the observer is
at an angle $\zeta$ relative to the spin axis.  The location of an
X-ray-emitting surface element on a neutron star relative to the
observer as a function of the time-varying rotational phase of the
pulsar, $\phi(t)$, is then defined by the angle $\theta$ between the
normal to the surface and the line of sight:
\begin{equation}
\cos\theta(t)=\sin \alpha \sin \zeta \cos \phi (t) + \cos \alpha \cos \zeta
\end{equation}
As the pulsar rotates, the varying projection of the emission area(s)
causes flux variations (i.e. pulsations), with shape and amplitude
determined in great part by the combination of $\alpha$ and
$\zeta$. Note that by convention $\alpha$ is reckoned from the spin
pole towards the equator.  The observed flux per unit energy from an
emitting region is given by
\begin{equation}
F(E)=I(E){\rm d}\Omega
\end{equation}
where $I(E)$ is the intensity of the radiation as measured by a
distant observer and ${\rm d}\Omega$ is the apparent solid angle
subtended by the emission region.  Transforming both quantities to the
rest frame of the emitting region yields
\begin{equation}
F(E)=(1-R_S/R_{NS})^{1/2} I'(E',\theta')\cos\theta\frac{{\rm d} \cos\theta}{{\rm d} \cos\psi} \frac{{\rm d}S'}{D^2}
\end{equation}
where the primed quantities are measured in the NS surface rest frame
\citep{Pou03}, with ${\rm d}S\cos\theta={\rm
  d}S'\cos\theta'$. $R_S\equiv 2GM/c^2$ is the Schwarzschild radius,
$I'(E',\theta')$ is the emergent intensity, which may be a function of
emission angle in addition to energy, ${\rm d}S'$ is the emission area
and $D$ is the distance.

A photon emitted at an angle $\theta>0$ with respect to the local
radial direction follows a curved trajectory and is observed at
infinity at an angle $\psi>\theta$. The relation between these two
angles is given by \citep{Pech83}:
\begin{equation}
\psi= \int_{R}^{\infty}\frac{{\rm d}r}{r^2}\left[\frac{1}{b^2}-\frac{1}{r^2}\left(1-\frac{R_S}{r}\right)\right]^{-1/2}
\end{equation}
where 
\begin{equation}
b=\frac{R_{NS}}{\sqrt{1-R_S/R_{NS}}}\sin\theta
\end{equation}
is the impact parameter at infinity of a photon emitted from radius
$R_{NS}$ at an angle $\theta$.  For most real-world applications, including
the analysis presented herein, a simplified approximate relation
between $\psi$ and $\theta$ \citep{Belo02} can be used:
\begin{equation}
\cos\psi\approx\frac{\cos\theta-R_S/R_{NS}}{1-R_S/R_{NS}}
\end{equation}
which is valid for $R_{NS} > 2R_S$.  This approximation greatly boosts the
computational speed of the model while still maintaining a high degree
of accuracy ($\lesssim$3\% error for $R \ge 3R_S$), allowing a thorough
exploration of the model phase space and implementation of more
complex emission regions. Owing to the relatively long spin period,
special relativistic effects, such as Doppler boosting and aberration,
as well as travel time differences are completely negligible
\citep{Pou03}.

The total observed flux for a given rotational phase is found by
relating $\phi$ and $\theta$ for a given emitting region through $\psi$ via
equations (1) and (4), using the desired $I'(E',\theta')$ in equation
(3), and summing the computed flux from all surface elements.  This
approach can be used to construct an arbitrary emission region on the
NS surface by considering any number and arrangement of surface
elements, provided they are sufficiently point-like so as not to
introduce significant errors in the model \citep[see, e.g.,][]{Tur13}.

%
%
\begin{figure}[!t]
\begin{center}
\includegraphics[width=0.45\textwidth]{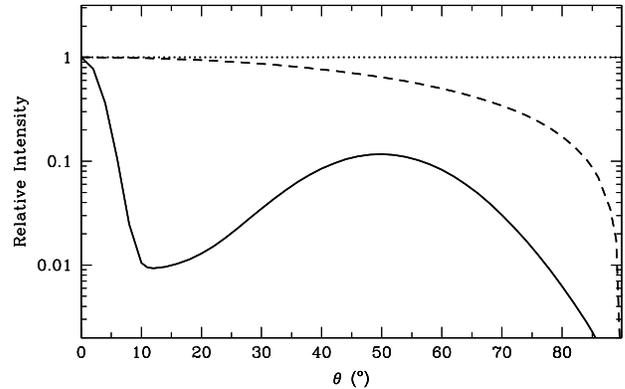}
\end{center}
\caption{The relative intensities as a function of angle with respect
  to the surface normal ($\theta$) for the three emission patterns
  considered in this analysis: isotropic (dotted), cosine
  beaming (dotted), and ``pencil plus fan'' beaming (solid).}
\end{figure}

\subsection{Surface Emission Model}
The surface composition of PSR J1852+0040, and CCOs in general, is
highly uncertain as there are multiple plausible possibilities. For
instance, a light element atmosphere may, in principle, build up due
to spallation of fallback material after the supernova explosion. Due
to gravitational sedimentation, as the lightest elements, hydrogen or
helium are expected to surface rapidly and thus dominate the surface
emission if a layer thicker than $\sim$1 cm accumulates
\citep{Chang04}. On the other hand, if no accretion takes place or if
thermonuclear reactions occur after accretion, a mid-Z element
\citep[see, e.g.,][]{Mori07,Ho09,Chang10} or iron atmosphere may be
present.  The same is likely to be the case if any pulsar wind outflow
is active that prevents accretion of material from the remnant.  It is
also quite possible that the stellar surface is devoid of an
atmospheric layer, in which case the emission from the condensed
neutron star surface may be reasonably well approximated by a
blackbody \citep{Pot12}.

Although based on the spin-down measurement the implied surface field
at the magnetic equator of PSR J1852+0040 is $3.1\times10^{10}$ G, the
strong X-ray pulsations may be a manifestation of a substantially stronger
field at the location of the hot regions, although the value cannot be
easily determined as there are no obvious absorption features (see
\S5.1).  This poses an additional difficulty in choosing the
appropriate surface radiation model, considering that the emission
characteristics of neutron star atmospheres can differ markedly
between $\sim$$10^{10}$ G \citep{Sule12}, $\sim$$10^{12}$ G
\citep{Pavlov94,Zavlin95}, and $\gtrsim$$10^{14}$ G
\citep[e.g.,][]{vanAd06}. As a consequence, any inferences drawn from
modeling the thermal emission are likely to be dependent on the true
surface magnetic field strength and its orientation relative to the
surface.

For strongly magnetized atmospheres ($\gtrsim$$10^{12}$ G), a narrow
``pencil'' beam along the direction of the magnetic field can also
appear in instances when the observer’s line of sight crosses the
magnetic field lines, as well as a broad ``fan'' beam with peak
intensity at intermediate angles with respect to the surface normal
\citep[see, e.g.,][]{Pavlov94,Zavlin95}. For weakly magnetic models
($\lesssim$$10^{10}$ G), the emergent intensity declines with
increasing angle with respect to the surface normal, resulting in a
limb-darkening effect \citep{Rom87,Zavlin96}.  For atmospheres with
$\sim$$10^{10-11}$ G, the emission is strongly beamed at photon
energies coincident with the harmonics of the cyclotron resonance
frequency, with the strongest beaming occurring at the fundamental
frequency and becoming progressively weaker with increasing harmonic
number. Away from the cyclotron absorption lines, the emission
generally declines with increasing angle away from the surface normal
\citep[see][in particular their Figure 7]{Sul10}.

Based on this information, to account for the variety of possible
angle-dependent intensity patterns of the thermal radiation from PSR
J1852+0040, I consider three possibilities: (i) a standard
isotropically emitting Planck spectrum; (ii) an emission model with a
cosine dependence of the intensity as a function of emission angle
relative to the surface normal as a proxy for a weakly magnetic
neutron star atmosphere, including a $\sim$$10^{10}$ G light-element
atmosphere at photon energies away from the cyclotron
harmonics\footnote{As shown in \S5, the phase-resolved spectra of PSR
  J1852+0040 exhibit no strong cyclotron harmonics so this assumption
  is appropriate.}; and (iii) a ``pencil plus fan'' beam pattern
characteristic of strongly magnetic atmospheres for the case of a
magnetic field perpendicular to the surface\footnote{As an
  approximation, I adopt the H atmosphere beaming pattern from
  \citet{Pavlov94} for $T_{\rm eff}=10^6$ K and $B=4.7\times10^{12}$ G
  at a photon energy of 2.28 keV.}. The three emission patterns are
illustrated in Figure 1. Although none of the models account for the
energy-dependence of the emergent intensity patterns of realistic
atmosphere models \citep[][]{Rom87,Shib92,Pavlov94,Zavlin95}, for the
purposes of this analysis the latter two provide an adequate
representation of the angular dependence (i.e.~``beaming'') produced
by a variety neutron star atmospheres, while being substantially less
computationally demanding than the full models. Therefore, although
the exact values of the parameters derived throughout this analysis
may not correspond to the actual values, the general conclusions
regarding the emission properties and heat distribution of the stellar
surface should be robust.

Neutron star atmospheres have the general property of producing
continuum radiation with peak intensities at higher energies relative
to a Planck spectrum for the same effective temperature
\citep{Rom87,Shib92,Ho01}.  As a consequence, when applied to thermal
spectra they tend to yield lower temperatures and hence larger
inferred emitting areas compared to a blackbody model. To account for
this property while minimizing the additional computational cost, for
the cosine beaming model, I use the empirical relation for
non-magnetic H atmospheres given by McClintock et al.~(2004; see in
particular their equations A17 and A18).  For the pencil plus fan beam
model, I implement a ``color correction'', obtained as follows.  The
spectrum of PSR J1852+0040 was fitted seperately with a blackbody
model and a magnetic {\tt nsa} atmosphere with $1\times10^{12}$ G. The
ratio of the derived emitting areas from the two models as a function
of temperature was used as a multiplicative factor to correct the flux
normalization in the pulse profile fits in order to obtain emitting
areas comparable to those of an actual magnetic atmosphere model.

%
%
\begin{figure}[!t]
\begin{center}
\includegraphics[width=0.45\textwidth]{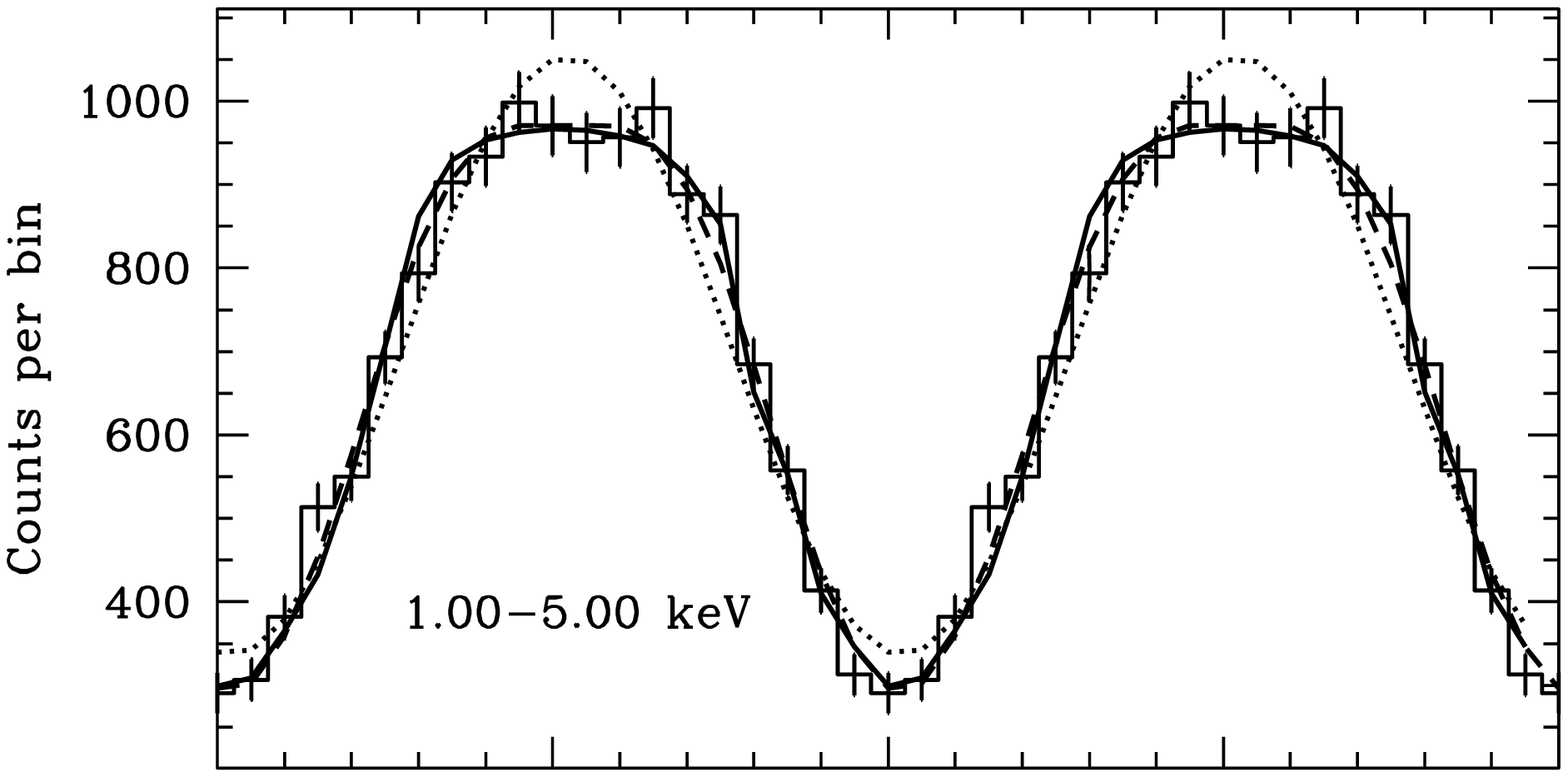}
\includegraphics[width=0.45\textwidth]{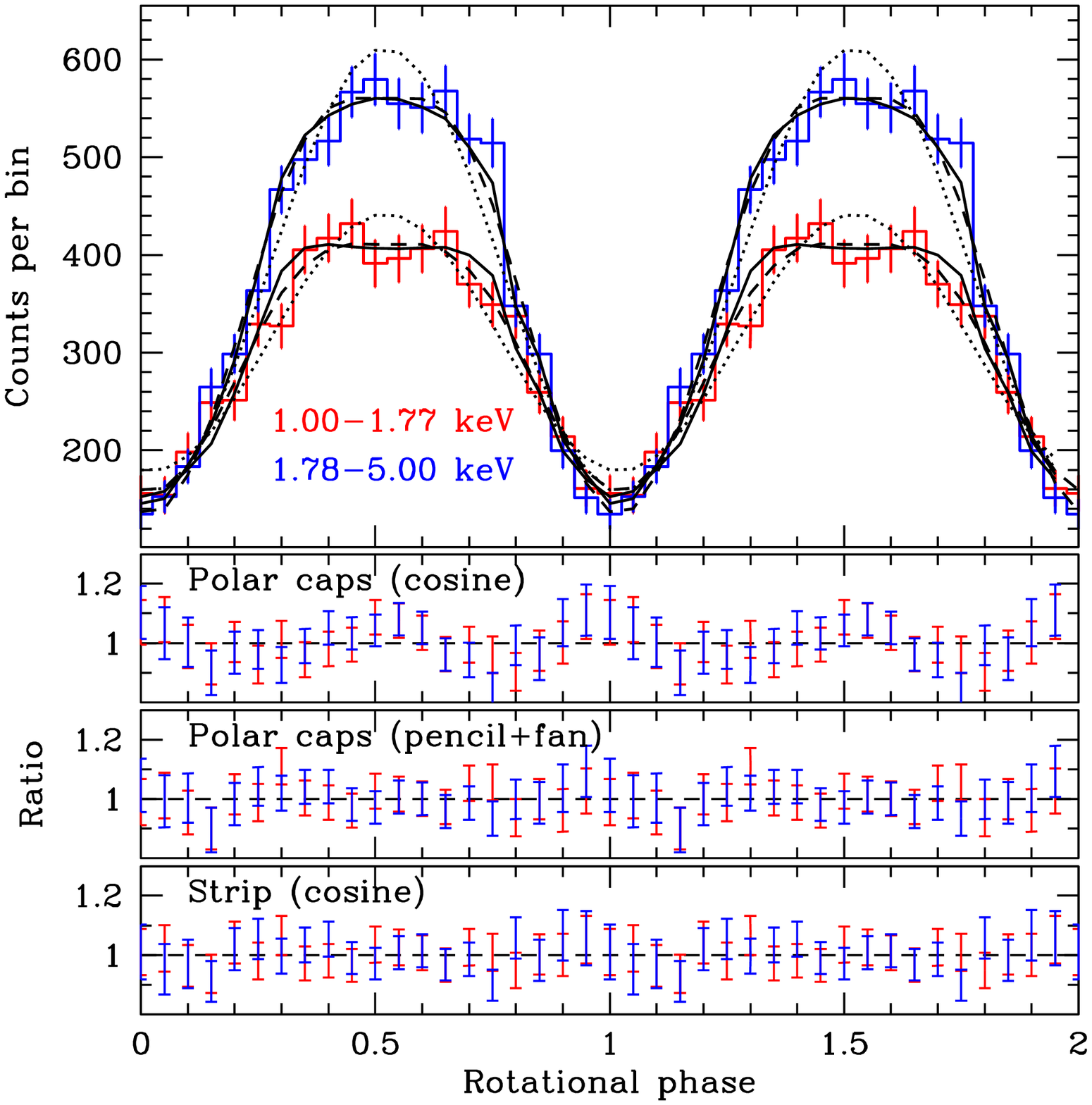}
\end{center}
\caption{(\textit{Top}) \textit{XMM-Newton} EPIC pn
  background-subtracted pulse profile of PSR J1852+0040 in the $1-5$
  keV interval. (\textit{Middle}) Pulse profiles in the $1-1.77$ and
  $1.78-5$ keV bands. The solid lines show the best fit for a rotating
  neutron star with two circular, antipodal hot spots for a ``pencil
  plus fan'' beam pattern. The dashed lines represent the best fit
  models of a rotating NS with a longitudinally extended hot strip
  across the surface for a cosine beaming emission model. The dotted
  lines correspond to the best fit for a rotating neutron star with
  two circular, non-antipodal hot spots for a cosine beaming model.
  The cosine and isotropic models produce virtually identical pulse
  profiles and residuals for both cases so only the former is
  shown. In all cases, a neutron star with $M=1.4$ M$_{\odot}$ and
  $R_{NS}=12$ km is assumed. The bottom three panels show the best fit
  residuals for the circular polar cap and strip models. Two
  rotational cycles are shown for clarity.}
\end{figure}

\subsection{Emission Region Geometry}
The spectra and pulse profiles accumulated from PSR J1852+0040 imply
one or more multi-temperature hot emission regions that are
significantly smaller than the full neutron star surface. For pulsar
in general, the location and geometry of the heated regions is
determined by the magnetic field structure at or beneath the stellar
surface
\citep[e.g.,][]{Heyl98,Heyl01,Pot01,Gep99,Gep06,Perez06,Pons09},
meaning that the surface emission can serve as a valuable tracer of
the field topology.  Previous studies by \citet{Shab12} and
\citet{Per13} have attempted to reproduce the observed pulse
properties of PSR J1852+0040 by computing the expected surface heat
signature of various assumed magnetic field configurations. However,
for the temperature distributions and emission models considered in
these investigations, the broad pulse shape and the large pulse
amplitude could not be simultaneously accounted for, hinting at a
strongly anisotropic emission pattern and/or a non-standard
arrangement of magnetic fields.  Herein, rather than start from an
assumed magnetic field configuration, I adopt the converse approach
and aim to deduce the surface emission properties and magnetic field
topology based on the heat distributions that can reproduce the
phenomenology of PSR J1852+0040 by fitting the synthetic pulse
profiles directly to the X-ray data.

For many thermally-emitting pulsars, a pair of circular hot spots,
presumably corresponding to the pulsar magnetic polar caps, provides
an adequate description of the observed thermal X-ray pulse
profile. Based on this, I consider antipodal as well as non-antipodal
polar caps (arising, for instance, due to an offset dipole), following
both the treatment of point-like hot spots presented in \citet{Belo02}
and of extended circular polar caps described in \citet{Got10} and
\citet{Tur13}.

The unusual pulse morphology offers qualitative insight regarding the
possible atmosphere emission pattern as well as the heat distribution
on the stellar surface.  In particular, the broad and effectively flat
pulse peak requires that the flux from the star appear essentially
unchanged to the observer for $\sim$20--30\% of the rotation
period. This can be produced by either a strongly anisotropic emission
pattern or a region on the surface that is elongated in the direction
of rotation ($\phi$). To explore the latter possibility, I focus on
regions on the surface that are much more extended in longitude than
in latitude. The simplest way to describe such a high aspect ratio
region using $\alpha$ and $\phi$ is to consider a strip of emission at
constant latitude, which can be parameterized by angular extents in
longitude and latitude ($\Delta\phi$ and $\Delta\alpha$,
respectively), and the values of $\alpha_o$ and $\phi_o$ of the
geometric center of the emitting region. For such a longitudinal
strip, the area is obtained by computing the integral of the region on
a sphere
\begin{eqnarray}
A_{\rm strip} &=& R_{NS}^2 \int_{\phi_o-\Delta\phi}^{\phi_o+\Delta\phi} \int_{\alpha_o-\Delta\alpha}^{\alpha_o+\Delta\alpha} \, \sin \alpha \,  \mathrm{d}\alpha  \, \mathrm{d}\phi \nonumber \\ 
 &=& 2R_{NS}^2\Delta\phi_o[\cos(\alpha_o-\Delta\alpha)-\cos(\alpha_o+\Delta\alpha)] 
\end{eqnarray}
This heat distribution can be easily modeled using Equation 3 by
dividing the emission region into a grid consisting of smaller surface
elements, each with an area defined by Equation 7.  Although as
defined, the rectangular shape of the strips is obviously not natural,
given the available photon statistics such a geometry is
indistinguishable from a more plausible one, such as an elliptical
region or a strip with rounded corners or semicircular end caps.
Moreover, the computational speed afforded by this simple
parametrization allows a thorough exploration of the model phase space
to identify the general type of heat distributions that can reproduce
the data.

%
%
\begin{figure}[!t]
\begin{center}
\includegraphics[width=0.45\textwidth]{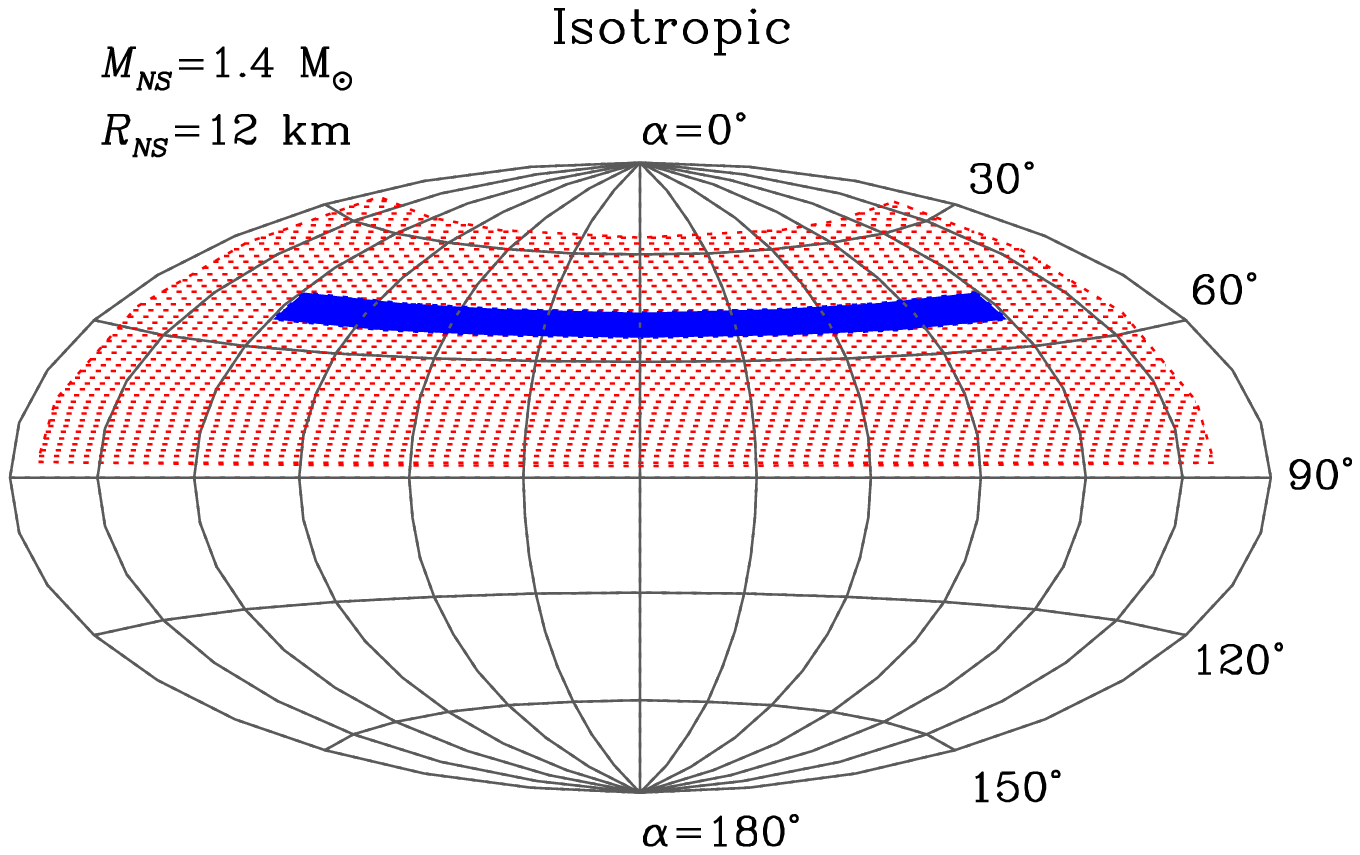}\\
\includegraphics[width=0.45\textwidth]{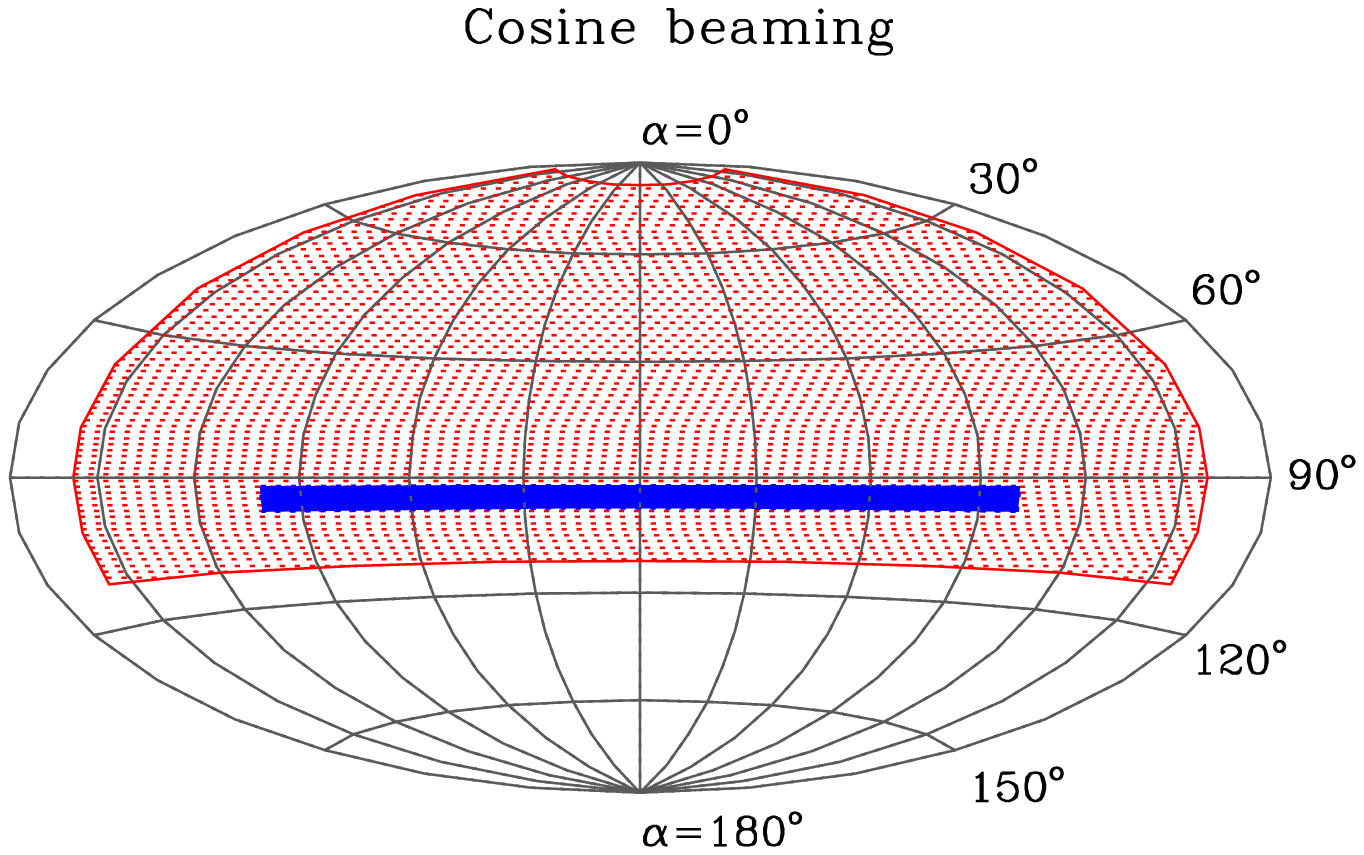}\\
\includegraphics[width=0.45\textwidth]{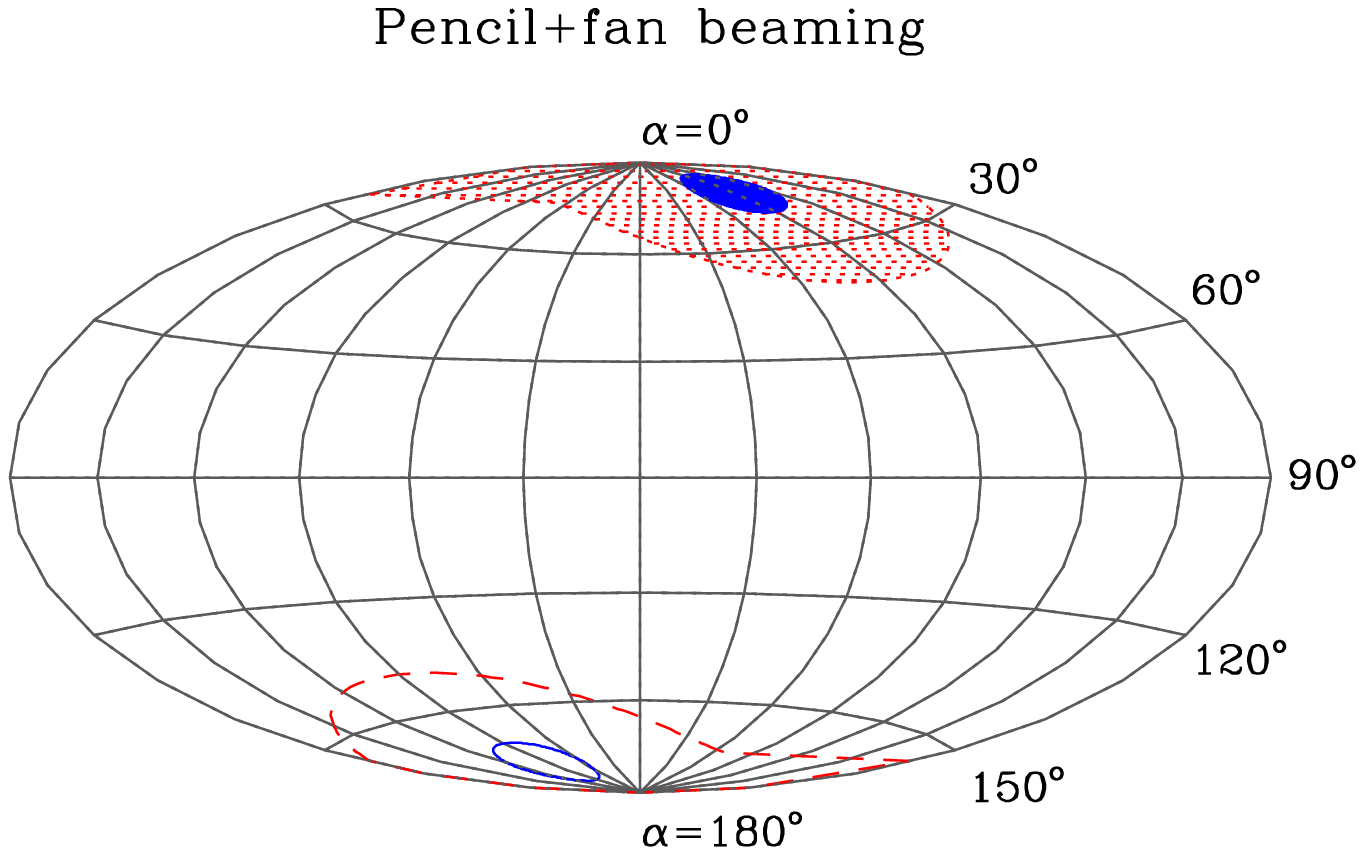}
\end{center}
\caption{Hammer-Aitoff equal area projection of the surface of PSR
  J1852+0040 showing the likely temperature distribution inferred from
  the pulse profile modeling for an assumed neutron star radius of 12
  km and mass $1.4$ M$_{\odot}$. The hot and warm regions are shown in
  blue and red, respectively. The isotropic, cosine, and pencil plus
  fan beam emission pattern results are shown from top to bottom,
  respectively. For the last case, the secondary polar cap does not
  contribute to the observed emission so only its outline is shown for
  reference.}
\end{figure}

\section{Pulse Profile Modeling Results}
To enable a direct comparison with observations, the synthetic pulse
profiles generated using the model described above were first
convolved with the EPIC pn detector response. As a consequence of the
high hydrogen column density along the line of sight towards the
pulsar, little useful spectral information is available below $\sim$1
keV. Based on this, I chose two energy bands which allows some
sensitivity to the spectral shape of the radiation.  The fits were
performed simultaneously in two energy bands, 1.0--1.77 and 1.78--5.0
keV, in which the warm and hot thermal components dominate,
respectively. Cooler surface emission from the rest of the
  neutron star is likely negligible above $\sim$1 keV so it is not
  modeled.  Throughout the analysis, I assume a neutron star with
$M=1.4$ M$_{\odot}$ and $R_{NS}=12$ km at a distance of $D=7.1$ kpc
and $N_{\rm H}=1.52\times10^{22}$ cm$^{-2}$ based on \citet{Gia09}.
To assess the dependence of the results on the highly uncertain
neutron star compactness, the analysis was repeated for other values
of $R_{NS}$ in the range $9-15$ km.

In the formal fits to the folded light curves I consider the allowed
range of values for $\alpha$ and $\zeta$ ($0^{\circ}-180^{\circ}$, and
the range of acceptable emission region areas and temperatures as
deduced from spectral fits. In the case of the circular cap model, I
also consider the radius of each hot spot, while in the longitudinal
strip model, the angular extents in longitude and latitude,
$\Delta\phi$ and $\Delta\alpha$ are additional free parameters.
Constraints on these parameters were derived via Monte Carlo
simulations of $5\times 10^3$ realizations for each combination of
stellar mass, radius and one of the three emission models described in
\S3.2.  In the pulse profile fits, the emission regions were
adaptively resized based on the input values of the angular extent of
the entire region in each direction.  The number of surface elements
($90$ and $45$ in the $\phi$ and $\alpha$ directions, respectively)
was chosen so as to ensure that the size of each is effectively
point-like, which is the case for angular extents
$\lesssim$$5^{\circ}$ \citep[see][]{Tur13}. For both the polar cap and
strip geometries, the hot and warm strips were allowed to intersect
such that in the overlap region the emission is solely due to the hot
region.

%
%
\begin{figure}[!t]
\begin{center}
\includegraphics[width=0.45\textwidth]{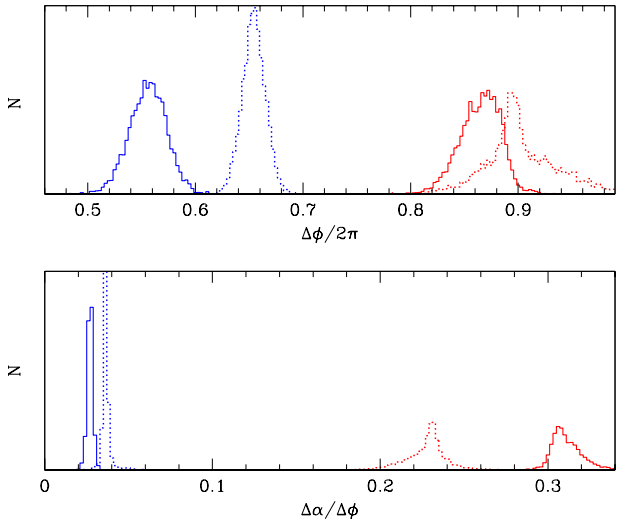}
\end{center}
\caption{Summary of results of the pulse profile fits of PSR
  J1852+0040 derived from Monte Carlo simulations, assuming a neutron
  star with $M=1.4$ M$_{\odot}$ and $R_{NS}=12$ km. The solid lines
  and dotted lines correspond to the beamed and isotropic emission
  models, respectively.  (\textit{Top}) The fraction of the stellar
  circumference subtended by the hot (blue) and warm (red) emission
  regions in longitude at the latitude of the centroid of the strip,
  $\Delta\phi/2\pi$ (\textit{Bottom}) Aspect ratio
  ($\Delta\alpha/\Delta\phi$) of the angular extent of the X-ray
  emitting regions in latitude ($\Delta\alpha$) and longitude
  ($\Delta\phi$).}
\end{figure}

%
%
\begin{figure}[!t]
\begin{center}
\includegraphics[width=0.45\textwidth]{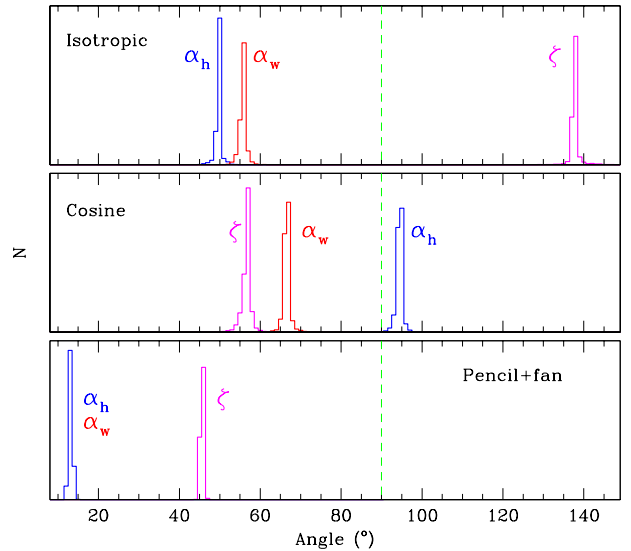}
\end{center}
\caption{The constraints on the angle $\alpha$ of the centroid of the
  hot (\textit{red}) and warm (\textit{blue}) regions and the viewing
  angle $\zeta$ (\textit{magenta}). The results for the isotropic,
  cosine, and pencil plus fan beam patterns are shown from top to
  bottom, respectively. For the latter, the hot and worm regions are
  co-located so $\alpha$ is identical for both. In all cases, a
  symmetric set of model solutions is obtained if the three angles are
  mirrored about the stellar equator, corresponding to 90$^{\circ}$ in
  this plot (\textit{green dashed line}).}
\end{figure}

Figure 2 shows the best fits to the \textit{XMM-Newton} EPIC pn pulse
profile of PSR J1852+0040 with a model of a NS with a longitudinal
heated strip, as well as a pair of circular, non-antipodal hot
caps. For both the polar cap and strip geometries, the isotropic
blackbody and cosine beaming models yield virtually identical best-fit
model pulse profiles.  For these emission patterns, it is apparent
from the systematic residuals that the conventional polar cap model
has difficulty simultaneously accounting for the wide and flattened
peak and the narrow trough, even if the assumption of antipodal hot
spots is relaxed. The best fit results in $\chi_{\nu}=1.48$
  for 31 degrees of freedom. Therefore, if the surface emission
exhibits an isotropic emission pattern or limb-darkening typical of
weakly magnetized atmospheres, the heated regions on the surface of
PSR J1852+0040 cannot be circular caps. In contrast, the ``pencil plus
fan'' beam model can easily reproduce the pulse shape with the
conventional antipodal polar cap model, yielding
  $\chi_{\nu}=0.90$ for 31 degrees of freedom. This is possible
because of the fan component of the beam, which peaks at intermediate
angles with respect to the surface normal (Figure 1) and is thus able to
compensate for the decline in the projected area of the polar cap as
the star rotates, resulting in a broad, flat-topped pulse. The
  best fit redshift-corrected temperatures are
  $T_h=(7.4\pm0.6)\times10^6$ K and $T_w=(3.2\pm0.6)\times10^6$ K,
  with the corresponding polar cap emission radii of
  $R_{h}=1.5\pm0.6$ km and $R_{w}=6.0\pm2.5$ km. The quoted
  uncertainties are at a 1$\sigma$ confidence level.

The elongated strip configuration is able to reproduce the
  pulse shape for both the isotropic and cosine beaming models,
  resulting in best fits with $\chi_{\nu}=0.95$ for 29 degrees of
  freedom in both instances.  The isotropic emission model produces
  best fit tempeartures (as measured at the neutron star surface) of
  $T_h=(7.0\pm0.6)\times10^6$ K and $T_w=(3.3\pm0.8)\times10^6$ K. The
  half-widths of the hot and warm regions in the latitudinal direction
  are $R_{\alpha_h}=0.7\pm0.3$ km $R_{\alpha_w}=6.5\pm1.4$ km, and
  $R_{\phi_h}=18.9\pm4.9$ km $R_{\phi_w}=28.2\pm7.1$ km in the
  longitudinal direction.  For the cosine beaming model, the best fit
  parameters are $T_h=(3.2\pm0.5)\times10^6$ K and
  $T_w=(1.7\pm0.6)\times10^6$ K, $R_{\alpha_h}=0.56\pm0.11$ km and
  $R_{\alpha_w}=9.3\pm3.8$ km, and $R_{\phi_h}=20.9\pm6.3$ km and
  $R_{\phi_w}=24.2\pm8.4$ km.  Note that for all three emission
  models, the quoted values of $R_{h}$ and $R_{w}$ correspond to arc
  lengths on the stellar surface.

For the assumed $M=1.4$ M$_{\odot}$, the isotropic and cosine beaming
emission pattern models produce no acceptable solutions for
$R_{NS}\lesssim9$ km, while for the pencil plus fan beam, the same is
true for $R_{NS}\lesssim8.5$ km. In the case of the isotropic model,
this is expected since for more compact stars it is not possible to
produce the remarkably large observed pulse amplitude because of the
stronger gravitational bending of light effect, which acts to greatly
diminish the amplitude of rotation-induced modulations \citep[see,
  e.g.,][]{Psa00}.  For beamed emission, this effect is not as strong
and a much larger pulsed fractions can in principle be achieved for
the same set of model parameters because the anisotropic emission
pattern of the emergent intensity acts to counter the suppression of
pulsations caused by light bending \citep[see, e.g.,][for the case of
  the millisecond pulsar PSR J0030+0451, with a $\sim$70\% thermal
  pulsed fraction]{Bog09}. However, in the case of PSR J1852+0040 in
particular, the area of the warm emission region required to produce a
satisfactory fit exceeds the total surface area of a $9$ km and $8.5$
km neutron star, for the cosine and pencil plus fan beam patterns,
respectively.

Figure 3 illustrates the most probable geometric configurations of the
emission regions deduced from the pulse profile modeling for a neutron
star with radius of 12 km. Similar configurations are obtained for the
range of plausible neutron star radii considered in the
analysis. Figures 4 and 5 show summary plots of the various parameters
of the fit based on the array of Monte Carlo simulations.  Several
noteworthy features of the inferred emission regions are evident.  In
particular, for the isotropic and cosine beaming patterns, the results
favor emission regions that have substantial elongation in the
longitudinal direction, with aspect ratios ranging from 3:1 for the
warm emission region to nearly 100:1 for the hot component.  The
requirement for such extreme aspect ratios to reproduce the data
explains why the conventional polar cap model cannot fit the pulse
profile using these emission patterns. Even in the case of two polar
caps that are adjacent and aligned in the $\phi$ direction, it is only
possible to obtain an aspect ratio up to $\sim$2:1.

As evident from Figure 3, for the elongated strip geometry the hot
emission region tends to lie well away from the spin poles, which is a
necessary condition for producing the large pulse amplitude at higher
photon energies.  The warm region is substantially more extended in
both longitude and latitude, nearly wrapping around the star and
covering up to $\sim$50\% of the entire stellar surface (see, e.g.,
the top two maps in Figure 3).  In general, the fits favor co-located
hot and warm regions, especially a thin hot strip enveloped entirely
by a much larger warm region. This suggest that the thermal X-ray
radiation originates from a single contiguous multi-temperature
region.  It is possible that the strips are not at constant
latitude, but are instead inclined with respect to the spin
equator. However, accounting for this would require the introduction of
an additional free parameter, which, given the excellent fit of the
current model, is not warranted by the data. Moreover, any such
inclination is likely small (of order a few degrees) since a highly
inclined hot strip would not reproduce the observed flat pulse.

The best fit for the pencil plus fan beam model places the polar caps
near to the spin pole (which actually lies within the larger warm
region; see bottom of Figure 3) but the highly anisotropic beaming
pattern is still able to produce a large amplitude pulse.  Although
for the pencil plus fan beam model two identical, antipodal polar caps
were assumed, for the best fit geometry, the second polar cap resides
in the region not visible to the observer. As a result, identical
results are obtained with a single polar cap and the properties of the
second polar cap are poorly constrained.

For a given surface emission model and assumed stellar radius, the
geometry and the location of the emission regions are very tightly
constrained, owing to the unique morphology and large amplitude of the
X-ray pulsations.  It should be noted, however, if the possible values
of the neutron star mass and radius, uncertain surface composition,
and magnetic field strength are considered, the allowed range of the
free parameters become quite large.  In addition, although the X-ray
emitting areas are assumed to be at two discrete temperatures, in
reality, a smooth temperature gradient likely exists between the warm
and hot regions.

%
%
\begin{figure}[!t]
\begin{center}
\includegraphics[width=0.45\textwidth]{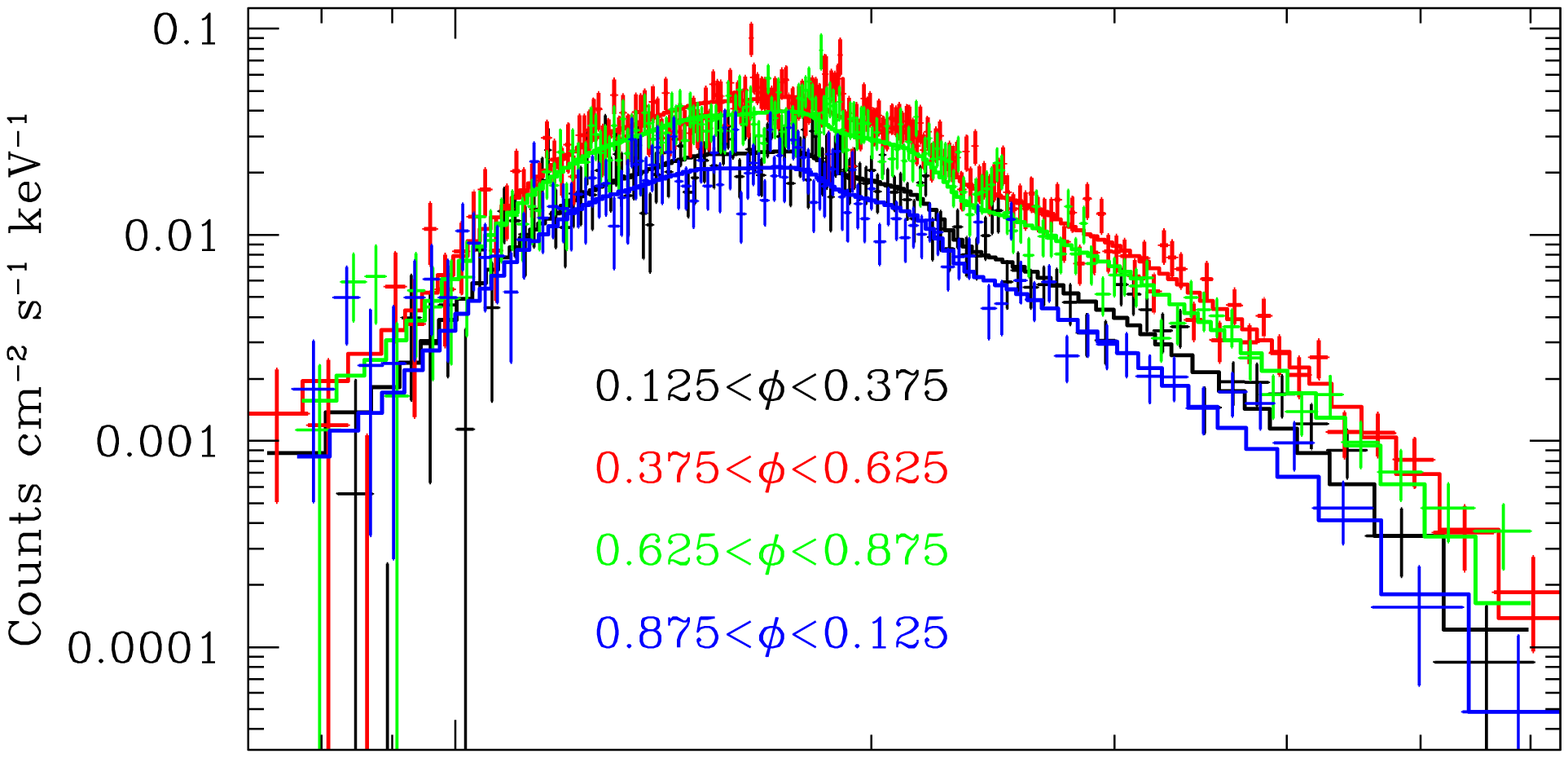}
\includegraphics[width=0.45\textwidth]{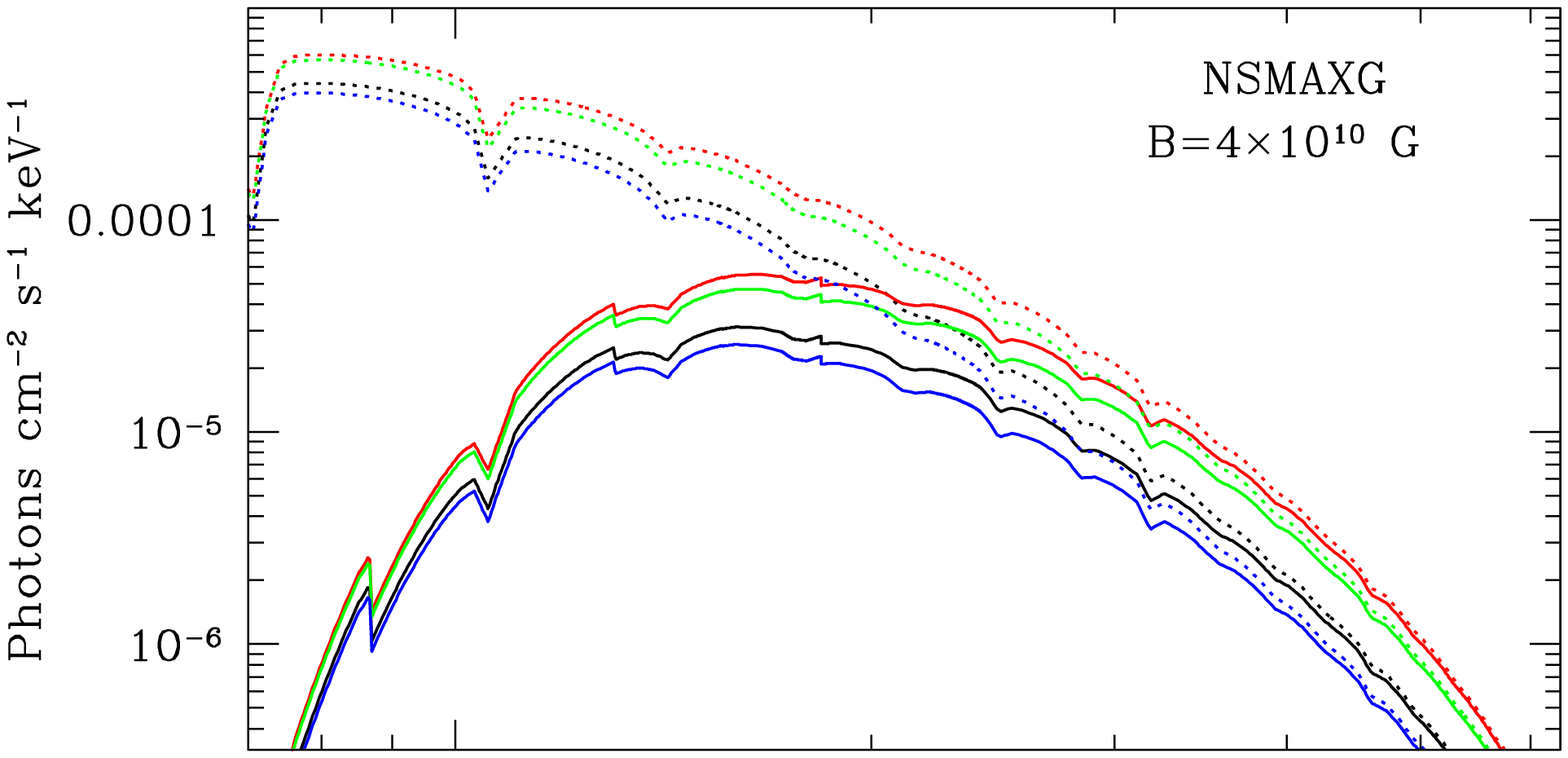}
\includegraphics[width=0.45\textwidth]{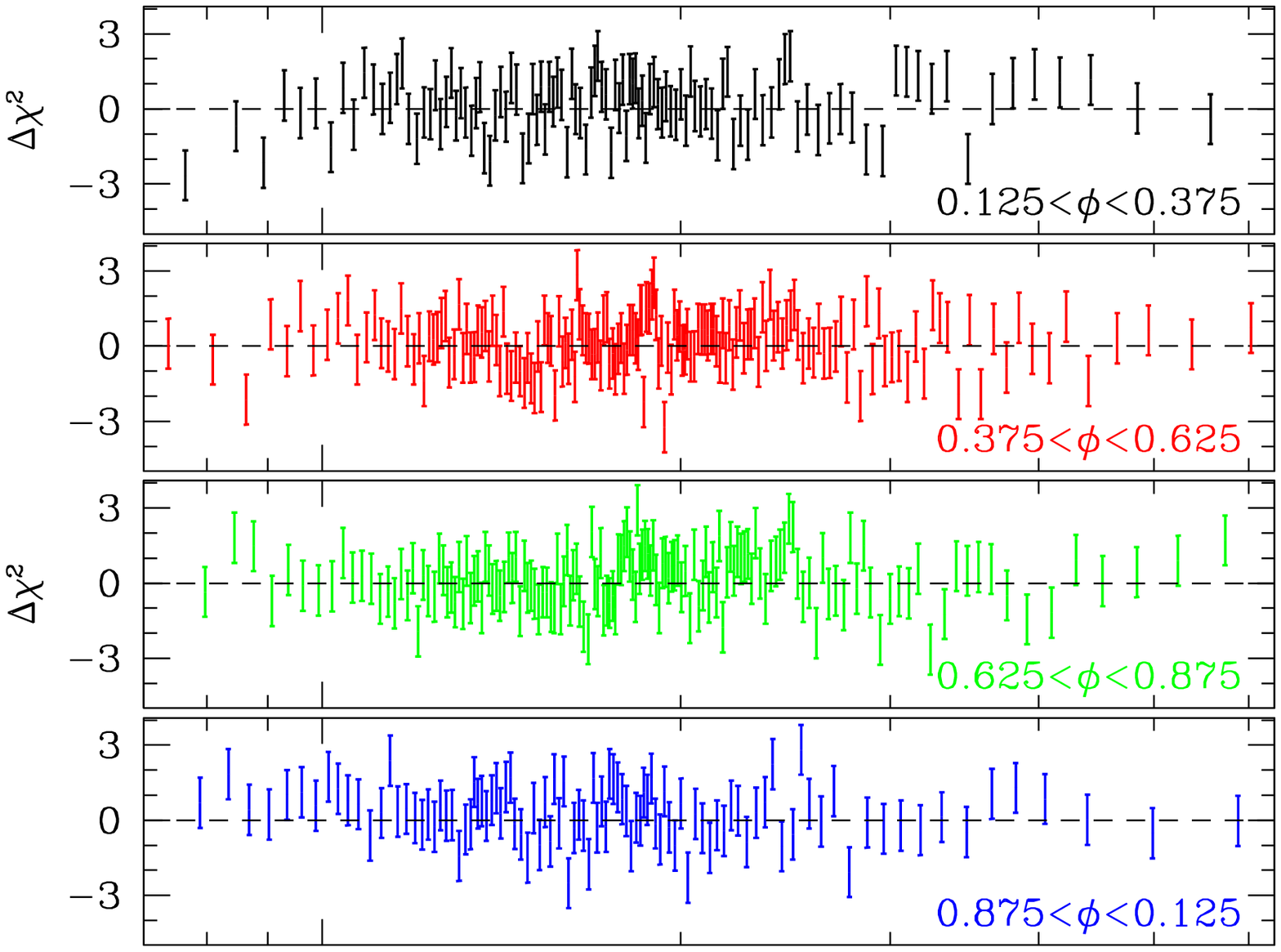}
\includegraphics[width=0.45\textwidth]{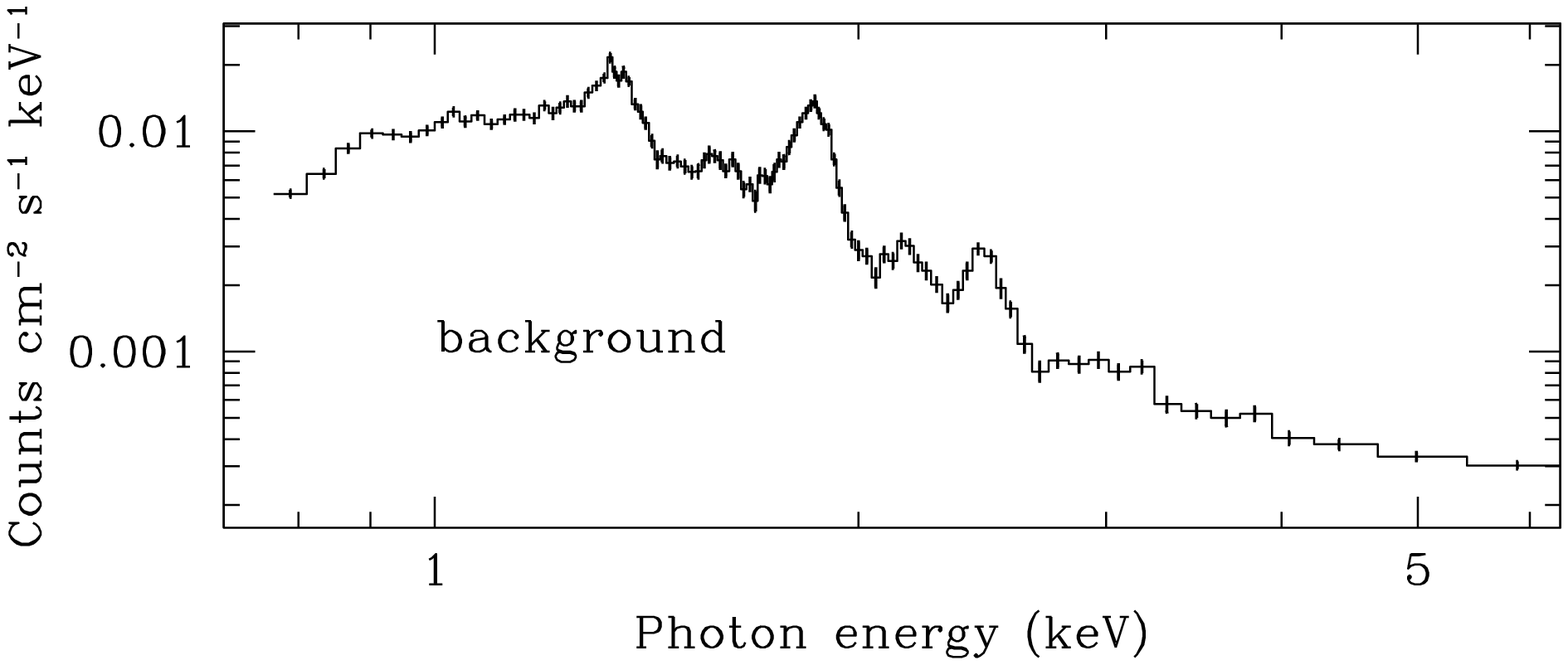}
\end{center}
\caption{(\textit{Top}) Pulse phase-resolved spectra of PSR J1852+0040
  fitted with a two-temperature $4\times10^{10}$ Gauss H atmosphere
  ({\tt nsmaxg} model number 1060). The phase intervals are based on
  the pulse profiles in Figure 2. (\textit{Second from top}) The best
  fit absorbed (solid) and unabsorbed (dotted) model spectrum.  The
  four panels show the best fit residuals expressed in terms of
  $\sigma$ with error bars of size one. (\textit{Bottom}) The
  background spectrum used in the spectroscopic analysis.}
\end{figure}

\section{Phase-resolved Spectroscopy}

\subsection{A Search for Narrow Spectral Features}
The available \textit{XMM-Newton} data provide a sufficient harvest of
source photons to allow an investigation of any phase-dependent
spectral features.  This is of particular relevance for CCOs given
that 1E1207.4--5209 at the center of the supernova remnant PKS
1209--51/52 shows two distinct harmonically-related features at $0.7$
and $1.4$ keV \citep{Big03,DeL04}, plus two features at $2.1$ and
$2.8$ keV whose existence is questionable \citep{San02,Mori05}. These
absorption exhibit remarkable variability as a function of spin
phase. The most plausible interpretation is that they arise
  due to resonant cyclotron absorption, with the $0.7$ keV feature
  corresponding to the fundamental frequency. If the absorption arises
  from electrons near the neutron star surface, the relation between
  the fundamental cyclotron energy (corrected for gravitational
  redshift) and magnetic field, $E_c=\hbar e B/mc=0.116(B/10^{10}~{\rm
    G})$ keV, implies a field strength of $8 \times 10^{10}$ G.  An
  alternative interpretation focuses on helium-like oxygen or neon in
  a magnetic field of $\sim$$10^{12}$ G \citep[see,
    e.g.,][]{Hailey02,Mori05}.

Motivated by this, I extracted phase-resolved spectra from the
archival data in Table 1 using four equal pulse phase intervals:
$0.125<\phi_1<0.375$, $0.375<\phi_2<0.625$, $0.625<\phi_3<0.8755$, and
$0.875<\phi_3<0.125$. As shown in Figure 2, as defined, phase zero
coincides with the pulse minimum. For each spectrum, the counts were
grouped so as to ensure at least 30 counts per energy bin. All four
spectra were fitted jointly in {\tt XSPEC} 12.7.1 \citep{Arnaud96}
with the same temperatures for all four phases but with independent
flux normalizations.

In \citet{Halpern10}, only a blackbody and non-magnetic {\tt nsa}
models were considered. However, in the standard vacuum dipole
radiation formalism the expected magnetic field strength at the
magnetic poles of PSR J1852+0040 is approximately $2B_{\rm surf}
\simeq 6.1\times10^{10}$ G. A more realistic treatment \citep[see,
  e.g.,][]{Spit06} yields $(4-5)\times10^{10}$ G, depending on the
magnetic inclination. Based on this, I consider a model that is more
appropriate for this pulsar -- a two-temperature ({\tt nsmaxg})
neutron star hydrogen atmosphere model with $B=4\times10^{10}$ G
\citep{Ho08}. Since the results from \S4 for the pencil plus fan beam
model suggest a substantially stronger field at the polar cap, I also
employ the {\tt nsa} model with $B=1\times10^{12}$ G
\citep{Pavlov94}. In both cases, the models have been computed for
$M=1.4$ M$_{\odot}$ and $R_{NS}=10$ km.

 Statistically acceptable, equally good fits are obtained for both
 models (see Table 2).  As evident from Figure 6, which shows the best
 {\tt nsmaxg} model fit, several narrow-band residuals are
 apparent. However, the energies of these features coincide with
 features seen in the background spectrum (bottom panel of Figure
 6). Based on this, I conclude that they arise due to imperfect
 background subtraction.  As noted previously, PSR J1852+0040 is
 situated in a relatively X-ray-bright supernova remnant with
 significant spatially-dependent variations in brightness and spectral
 shape, causing difficulty in obtaining a representative background.

 Aside from these features, no statistically significant
 phase-dependent residuals that could be plausibly associated with
 cyclotron absorption/scattering are seen in the spectrum of PSR
 J1852+0040. This is not surprising given that the weaker magnetic
 field derived from spin-down relative to 1E1207.4--5209 only produces
 weak features above $\sim$1 keV from the higher order cyclotron
 harmonics, as evident from the second panel from the top in Figure 6
 \citep[see also][]{Sul10}. The most prominent absorption features of
 the $B=4\times10^{10}$ G model, corresponding to the fundamental (at
 $\sim$0.35 keV) and first overtone ($\sim$0.7 keV) of the cyclotron
 resonance, would be severely attenuated by interstellar absorption, while
 the remaining features are too shallow to be identified in the
 present data given the limited energy resolution and insufficient
 photon statistics.  In the $\sim$$10^{12}$ G scenario, no narrow-band
 spectral features are expected in the energy range under
 consideration.

The absence of any phase-dependent absorption features intrinsic to
the pulsar in the spectrum indicates that the observed
rotation-induced flux variations are unlikely to be due to
phase-dependent resonant cyclotron scattering of the surface thermal
X-rays from a uniformly emitting neutron star, as recently proposed
for ordinary pulsars \citep[see, e.g.,][]{Kar12}.  Even in the case of
1E1207.4--5209, which unambiguously shows absorption features, the
underlying cause for the pulsations is likely the changing view of the
hot regions on the star due to rotation, with the resonant scattering
only enhancing the rotation-induced flux modulations rather than being
the sole cause.

\subsection{A Carbon Atmosphere?}
For the CCO in Cas A, the derived effective radii of the emission
region for H and He atmosphere models ($4-6$ km) are much smaller than
the expected NS radius. Based on this and the apparent lack of X-ray
pulsations, \citet{Ho09} have argued that Cas A needs to be covered by
a non-magnetic C atmosphere in order to produce an emission size
$R=15.6^{+1.3}_{-2.7}$ km, assuming $M=1.4$ M$_{\odot}$ and $D=3.4$
kpc, that is consistent with the theoretical prediction for the radii
of NSs. More recently, \citet{Klo13} have applied a C atmosphere to
the X-ray spectra of the CCO in the HESS J1731--347/G353.6--0.7
remnant obtaining good fits for plausible values of the neutron star
mass and radius as well.

In light of these results it is interesting to compare the C
atmosphere fits to the phase-resolved spectra of PSR J1852+0040. For
this purpose, I have applied the recently published {\tt carbatm}
model \citep{Klo13,Sule14} to the four phase-resolved spectra
described in \S5.1. Table 3 summarizes the best fit parameters for
fixed $D=7$ kpc and $M=1.4$ M$_{\odot}$ and three assumed stellar
radii: 9, 12, and 14 km.  It is apparent that for 9 and 12 km, the
inferred area at pulse maximum exceeds the total surface area of the
star. Even in the case of 14 km, the implied emission area is
equivalent to $\approx$95\% of the NS surface. However, although a
single-temperature C atmosphere model produces a statistically
acceptable fit with $R_{\rm eff} \approx R_{NS}$, this finding cannot
be reconciled with the strongly pulsed X-rays from PSR J1852+0040,
which indicate emission from a much smaller portion of the stellar
surface.

After an age of about 1000 yr, a NS should cool enough to allow a
light element atmosphere to accumulate \citep{Chang10}. Based on this,
as Cas A is only $\sim$330 yr old \citep{Fes06}, a C atmosphere may in
fact be present on its surface \citep[see, however,][]{Poss13}.  On
the other hand, the ages of PSR J1852+0040 and the CCO in G353.6--0.7
have been estimated to be $5400-7500$ yr \citep{Sun04} and
$\sim$27,000 yr \citep{Tian08}, respectively.  Therefore, there is no
reason to expect a C atmosphere to dominate the surface emission in
these older CCOs.  Combined with this theoretical argument, the incongruity
of the C atmosphere result with the strong X-ray pulsations from PSR
J1852+0040 suggests that caution should be exercised when applying
such spectroscopic models to other CCOs as it could lead to specious
conclusions. This is especially true in cases where no pulsations have
been detected, meaning that no information regarding the actual
surface heat distribution can be gained.

\begin{deluxetable}{lc}
\tablewidth{0pt}
\tablecaption{Hydrogen atmosphere spectral fits for PSR J1852+0040.}
\tablehead{
\colhead{Parameter\tablenotemark{a}} & \colhead{Value}}
\startdata
\multicolumn{2}{c}{{\tt nsmaxg}\tablenotemark{b} ($B=4\times10^{10}$ G)}\\
\hline
$N_{\rm H}$ ($10^{22}$) cm$^{-2}$   & $1.52$  \\
$T_{\rm eff,1}$ ($10^6$ K)   & $3.46^{+0.11}_{-0.07}$ \\
$T_{\rm eff,2}$ ($10^6$ K)   & $1.48^{+0.48}_{-0.35}$ \\
$R_{\rm eff}$ (km)\tablenotemark{c}   & $3.2^{+3.3}_{-2.9}$ \\
$R_{\rm eff}$ (km)\tablenotemark{c}   & $4.9^{+9.4}_{-4.9}$ \\
$\chi^2_{\nu}$/dof & $1.05/530$ \\
\hline
\multicolumn{2}{c}{{\tt nsa}\tablenotemark{b} ($B=1\times10^{12}$ G)}\\
\hline
$N_{\rm H}$ ($10^{22}$) cm$^{-2}$   & $1.52$  \\
$T_{\rm eff,1}$ ($10^6$ K)   & $8.27^{+0.29}_{-0.31}$ \\
$T_{\rm eff,2}$ ($10^6$ K)   & $3.07^{+0.24}_{-0.25}$ \\
$R_{\rm eff,1}$ (km)\tablenotemark{c}    & $0.34^{+0.13}_{-0.11}$ \\
$R_{\rm eff,2}$ (km)\tablenotemark{c}    & $3.5^{+1.8}_{-1.4}$ \\
$\chi^2_{\nu}$/dof & $1.06/530$ 
\enddata
\tablenotetext{a}{Quoted uncertainties are at a 1$\sigma$ confidence level for one interesting parameter.}
\tablenotetext{b}{A neutron star of mass 1.4 M$_{\odot}$ and radius 10 km is assumed for both models.}
\tablenotetext{c}{Redshift-corrected effective emitting radius assuming $D=7.1$ kpc.}
\label{hatmtable}							 
\end{deluxetable}

\begin{deluxetable}{lccc}
\tablewidth{0pt}
\tablecaption{Carbon atmosphere spectral fits for PSR J1852+0040.}
\tablehead{
 \colhead{} & \multicolumn{3}{c}{$R_{\rm NS}$ (km)\tablenotemark{a}} \\
\cline{2-4}
\colhead{Parameter\tablenotemark{b}} & \colhead{9 km} & \colhead{12 km} & \colhead{14 km}}
\startdata
$N_{\rm H}$ ($10^{22}$) cm$^{-2}$   & $1.55\pm0.04$  & $1.52\pm0.04$ & $1.50\pm0.04$ \\
$T_{\rm eff}$ ($10^6$ K)   & $2.11 \pm 0.05$ & $1.91\pm0.05$ & $1.84\pm0.05$ \\
$A_{\rm eff}/A_{\rm NS}$\tablenotemark{c}   & $1.61^{+0.29}_{-0.24}$  & $1.16^{+0.21}_{-0.18}$ & $0.95^{+0.18}_{-0.15}$\\
$\chi^2_{\nu}$/dof & $1.11/534$ & $1.11/534$ & $1.11/534$
\enddata
\tablenotetext{a}{A neutron star mass of 1.4 M$_{\odot}$ is assumed in all cases.}
\tablenotetext{b}{Quoted uncertainties are at a 1$\sigma$ confidence level for one interesting parameter.}
\tablenotetext{c}{Effective emitting area expressed as a fraction of the total NS surface area assuming $D=7.1$ kpc.}
\label{catmtable}							 
\end{deluxetable}

\section{Discussion}

\subsection{Comparison with Other CCOs}
The pulse profile shape of PSR J1852+0040, especially the very broad
pulse, differs substantially from other thermally-emitting neutron
stars, including other CCOs like PSR 0821--4300 in Puppis A,
1E1207.4--5209 in PKS 1209--51/52, suggesting substantial differences
in temperature distribution and/or viewing geometry.

\citet{Got10} conducted detailed modeling of the X-ray pulsations and
spectra of PSR J0821--4300, the CCO in the SNR Puppis A. The analysis
demonstrated that a pair of thermal, diametrically opposite hot spots
on the surface is able to fully account for the observed two-component
thermal spectrum and energy-dependent pulse profile, including the
remarkable $180^{\circ}$ phase reversal at $\approx$1.2 keV. However,
the phase reversal requires that the temperatures of the two emission
spots differ by a factor of two and their areas by a factor of
$\sim$20.  In contrast, the markedly non-sinusoidal pulse profile of
PSR J1852+0040 exhibits no energy dependent phase shift.  This could
indicate that, unlike PSR J0821--4300, the emission regions of
different temperatures are either co-located on the surface or their
centroids are effectively at the same longitude.  Alternatively, this
may be the direct result of a surface heat map comparable to PSR
J0821--4300 but with different combination of magnetic inclination and
viewing angle. For PSR J1852+0040, in the best-fit antipodal hot spot
model obtained with the pencil plus fan beam emission model (see
bottom panel of Figure 3), the secondary polar cap does not contribute
significantly to the observed emission, which when combined with the
severe interstellar absorption of emission below $\sim$1 keV, would
not produce a pulse phase reversal due to a much larger, cooler
antipodal cap.

The CCO 1E 1207.4--5209 exhibits much less pronounced X-ray
pulsations, reaching a maximum $\sim$14\% pulsed fraction in the
fundamental cyclotron absorption feature at $\sim$0.7 keV
\citep{deluca08}. Aside from the enhancement in pulsations at energies
coinciding with the absorption features, the low-amplitude and
approximately sinusoidal pulsations suggest emission from a
conventional hot-spot configuration. The evidence for a slight phase
shift of the pulsations at energies below $\sim$0.5 keV, could be
interpreted using the same heat distribution found for PSR J0821--4300
but with a different combination of $\alpha$ and $\zeta$.

\subsection{A Strongly Magnetized Hot Spot?}
\citet{Shab12} have attempted to account for the X-ray properties of
PSR J1852+0040 by analyzing the expected heat distribution and
resulting X-ray light curves of a neutron star with a weak centered
dipole plus a strong ($\sim$$10^{14}$ G) toroidal crustal magnetic
field. The resulting heat distribution, characterized by small hot
spots, is capable of achieving a high X-ray pulsed fraction \citep[see
  Figures 2, 4, and 5 in][]{Shab12} but not a broad pulse shape that
closely resembles that of PSR J1852+0040. A likely explanation for
this is the assumption of a toroidal field that is large everywhere
except near the magnetic polar caps. This results in a weak field
($\sim$$10^{10}$ G) at the polar caps, which (away from the lower
order cyclotron harmonics) emits an emission pattern that is
well-approximated by a cosine beaming function. As shown in \S4, such
an emission pattern cannot account for the observed pulsations for the
standard antipodal hot spot model.

The spin-down measurement of PSR J1852+0040 implies a magnetic field
at the magnetic poles of $\sim$$4\times10^{10}$ G (for $R=12$ km and
moment of inertia $I=10^{45}$ g cm$^{2}$). However, the excellent fits
to the pulse profile with the polar cap model for the pencil plus fan
beam emission model suggests that the magnetic field needs to be
substantially higher ($\gtrsim$$10^{12}$) G to produce such a highly
anisotropic emission pattern.  This is contradictory to the weak
surface dipole field implied by the measured spin-down.  One way to
accommodate both findings is to displace the dipole field in the axial
direction such that at the magnetic pole closer to the magnetic moment
the field is significantly stronger, while at large distances from the
stellar surface the field still appears weak ($\sim$$10^{10}$ G). In
this sense, the implied heat distribution would be very similar to
that inferred for PSR J0821--4300 in Puppis A (see Figure 7a). As
noted in \S6.1, in this case the markedly different pulse properties
between the two CCOs can then be easily accounted for by different
combinations of $\alpha$ and $\zeta$.

\subsection{An Extremely Offset Dipole?}
The surface emission maps deduced using the isotropic and cosine
beaming patterns are quite peculiar, as they imply a lack of
discernable polar caps and the absence of azimuthal symmetry in the
surface emission. This could, in principle, arise due to large
deviations from a conventional centered magnetic field model.
\citet{Per13} have investigated the surface temperature profiles for
young, strongly magnetized ($10^{13-15}$ G) neutron stars by
considering both purely poloidal and a mixture of poloidal and
toroidal components magnetic fields.  Surprisingly, this analysis
revealed that for $\sim$5 kyr-old neutron stars (comparable to the age
of PSR J1852+0040) with both $10^{14}$ G poloidal and $5\times10^{15}$
G toroidal fields, the highest surface temperature is situated not at
the magnetic poles but in circumferential bands at intermediate magnetic
colatitudes, reminiscent of the strips illustrated in Figure
3. However, a key feature of the X-ray-emitting regions shown in
Figure 3 is the azimuthal asymmetry, namely partial hot bands that do
not completely encircle the star.  Indeed, the inherently axisymmetric
magnetic field configurations assumed in \citep{Per13} and similar
studies \citep[e.g.,][]{Gep06} cannot simultaneously account for the
broad pulse, narrow trough, and anomalously high pulsed fraction if
blackbody or emission patterns characteristic of weakly magnetized
atmospheres are assumed.  In principle, the necessary heat asymmetry
can be achieved by displacing the magnetic moment from the center of
the star or introducing a strong quadrupole component (provided that
the associated sub-surface field is of sufficient strength to
preferentially channel the interior heat to only a fraction of the
surface; see \S6.3).  A large displacement (of order $R_{NS}$) in a
direction orthogonal to the dipole axis would cause the magnetic polar
caps to become greatly elongated (as illustrated in Figure 7b). This
strip may not in fact be contiguous, with a gap between the two
``polar strips'', but at the phase resolution afforded by the photon
statistics of the presently available data such a gap is not
discernable.

\subsection{Submerged Strong Magnetic Fields?}
As noted by \citet{Halpern10} and numerous subsequent works, the
existence of hot areas that are a fraction of the total surface
for CCOs is difficult to reconcile with an intrinsically weakly
magnetized neutron star (i.e.~an ``anti-magnetar'') as it requires a
mechanism to confine the heat to a small region.  For strong fields,
the heat conductivity is enhanced in the direction parallel to the
magnetic field, while it is reduced in the perpendicular direction
\citep[e.g.,][]{Heyl98,Heyl01,Pot01,Gep99,Gep06,Perez06,Pons09}.
Hence, PSR J1852+0040 needs to possess a much larger ``hidden''
magnetic field in the crust than the dipole field inferred from the
spin-down measurement from \citet{Halpern10}.  This field acts as an
insulator thus restricting the surface heat to a portion of the
surface.

One plausible way to simultaneously account for the low apparent field
as measured from spin-down and the strong sub-surface field required to
explain the highly non-uniform surface heat distribution is to
consider the fallback of the debris of the supernova explosion onto
the newborn neutron star.  In particular, shortly after the violent
explosion, the neutron star is believed to accrete material from the
reverse shock at a rate greatly exceeding the Eddington limit
\citep[e.g.,][]{Blondin86,Chev89, Houck91}.  This episode of so-called
``hyper-critical'' accretion could bury the magnetic field into the
crust of the nascent neutron star, resulting in an apparent surface
field substantially weaker relative to the internal ``hidden''
magnetic field \citep{Vig12,Ber13}.  A post-supernova accretion
episode of $10^{-4}-10^{-3}$ M$_{\odot}$ over a large region of the
surface is necessary to bury the magnetic field into the inner crust.
This burial process can, in principle, result in crustal magnetic
fields with $\sim$$10^{14}$ G, which in turn, produce high temperature
contrast across the stellar surface, while still maintaining a low
apparent surface field.

The details of the current magnetic field topology presumably depend
on the particular geometry of the supernova explosion ejecta, and may
be the product of non-uniform fallback and/or low accretion rate
\citep[see][]{Ber13}.  In this scenario, the peculiar heat
distributions of PSRs J1852+0040 and J0821--4300 may be the direct
result of the configuration of the fallback material.  Alternatively,
it is possible that the natal magnetic field of the neutron star
deviated significantly from a centered dipole field in the first
place, possibly due to an off-center explosion \citep{Bur96,Lai00},
and the fallback uniformly submerged the field while still preserving
the initial global configuration but with a much weaker surface
field.

%
%
\begin{figure}[!t]
\begin{center}
\includegraphics[width=0.4\textwidth]{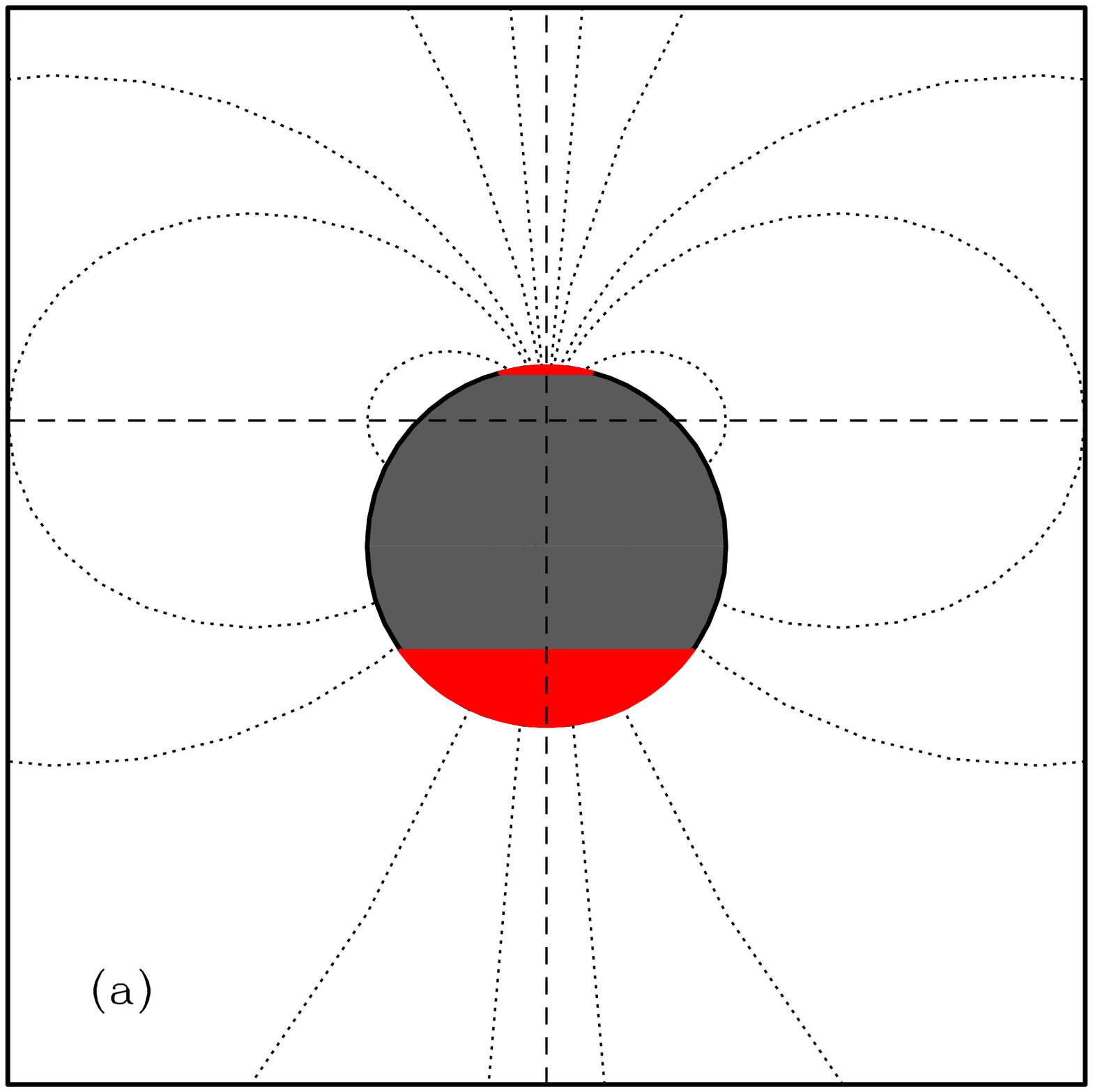}\\
\includegraphics[width=0.4\textwidth]{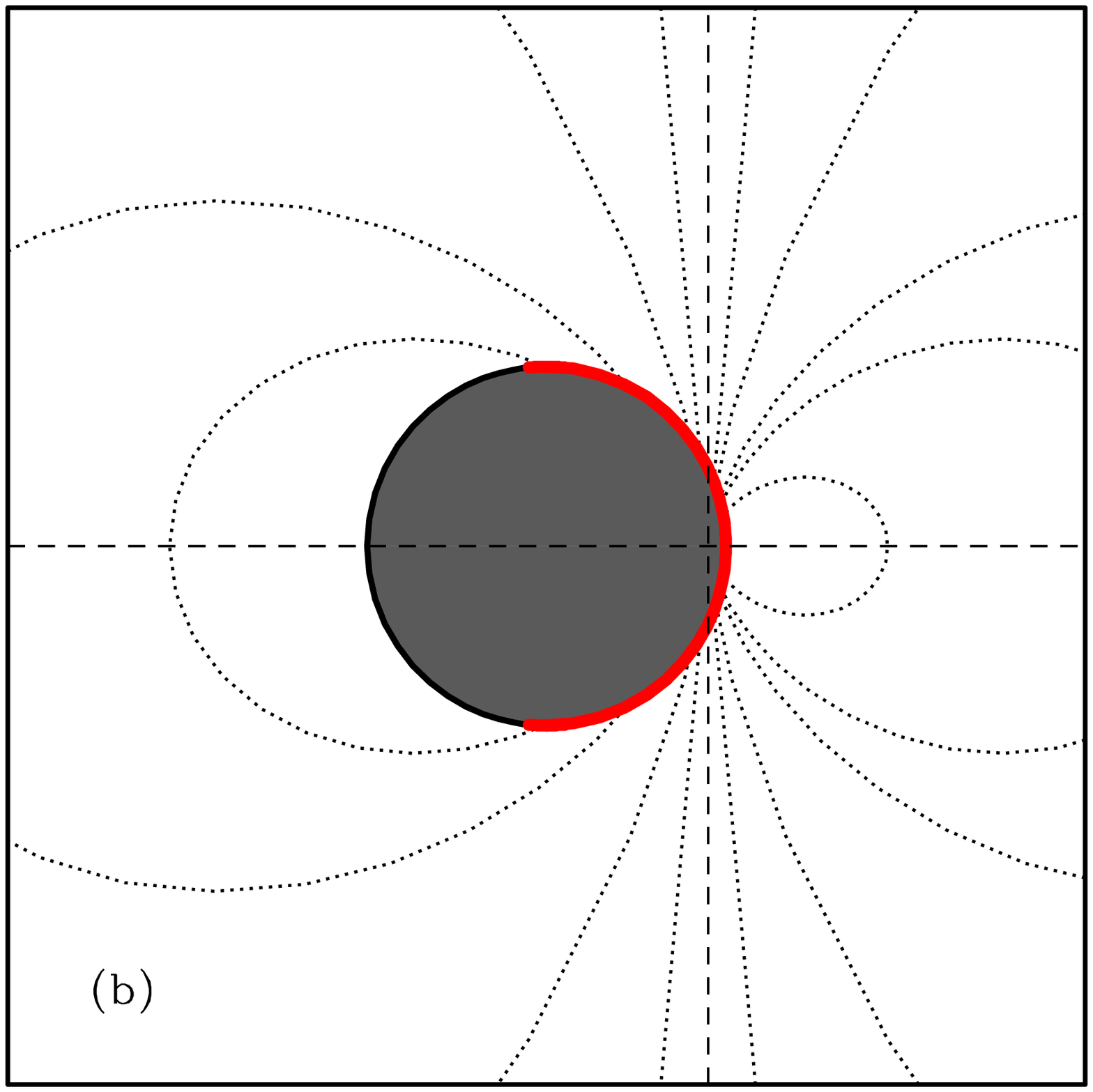}
\end{center}
\caption{Schematic illustration of the possible offset magnetic dipole
  field configurations for PSRs J1852+0040 discussed in the text: (a)
  A dipole offset in the direction of the magnetic axis and (b) a
  dipole offset perpendicular to the magnetic axis. Scenario (a) is
  also applicable to PSR J0821--4300.  The red shows the inferred
  X-ray emitting areas on the stellar surface. The dotted lines show
  the dipole magnetic field lines while the vertical and horizontal
  lines show the magnetic axis and equator, respectively.}
\end{figure} 

An alternative interpretation for the restricted heat on the surface
is on-going accretion of fall-back material at a sufficiently low rate
via a thin disk.  However, the steady spin-down over many years and
the lack of evidence for any long-term X-ray variability do not favor
this scenario.  Localized heating due to a return current, driven by
the rotation of the magnetized star, is also not likely as it requires
rotation-power to supply the energy, but the observed X-ray luminosity
greatly exceeds the spin-down luminosity of the pulsar.

\section{Conclusion}
I have presented modeling of the thermal X-ray pulsations from the
central compact object and 105-millisecond X-ray pulsar PSR J1852+0040
in the Galactic supernova remnant Kesteven 79. Unlike previous studies
\citep{Shab12,Per13}, the relatively simple models employed herein are
able to simultaneously account for both the X-ray pulse amplitude and
broad peak.  The unusual morphology of the pulse profile can be
reproduced with either: i) a conventional antipodal polar cap model
with a ``pencil plus fan'' beam intensity pattern; or ii) emission
regions on the stellar surface that are significantly elongated in
longitude (i.e., in the direction of rotation).  Although in the
analysis presented above only approximations to emission
models were considered for the sake of computational efficiency, the
main findings are likely to remain valid for a more realistic
treatment employing sophisticated atmosphere models of various surface
magnetic field strengths and chemical compositions.

Given that the observed thermal X-ray radiation from CCOs is due to
passive cooling, the inferred temperature distribution suggests highly
anisotropic heat conduction from the stellar interior. As posited by
several existing studies
\citep{Halpern10,Ho11,Vig12,Shab12,Ber13,Per13,Got13} if the heated
regions on the surface of PSR J1852+0040 are closely associated with
the magnetic field structure, strong magnetic fields beneath the
stellar surface are required to channel heat to a relatively small
portion of the star.  The constraints on the heat distribution of PSR
J1852+0040 presented herein further support the argument that rather
than being born with intrinsically weak fields, CCOs possess strong
``hidden'' magnetic fields that were buried due to rapid accretion of
fallback material shortly after the supernova explosion. This burial
hypothesis avoids the requirement for a strong external global dipole
magnetic field, which would manifest in the spin-down measurement.

An offset dipole can provide a plausible explanation for the two
surface temperature maps deduced in \S3 for PSR J1852+0040, while
still being consistent with the weak field inferred from the pulsar
spin-down.  In particular, for the ``pencil plus fan'' beaming model,
the implied strong surface field ($\gtrsim$$10^{12}$ G) needed to produce such
a highly anisotropic emission pattern can be explained by a magnetic moment
that is significantly displaced mostly along the axial direction of
the dipole (Figure 7a). This configuration can also account for the two
hot spots that differ greatly in size and temperature for the CCO in
Puppis A, PSR J0821--4300. The linear geometry of the heated regions
required for the isotropic and cosine beaming patterns can be
produced if the offset of the dipole is in a direction orthogonal
to the magnetic axis (Figure 7b). The large field displacements in
both cases are possibly a consequence of an off-center supernova
explosion.

A phase-resolved spectroscopic analysis reveals no phase-dependent
narrow-band features that could arise due to cyclotron
absorption/scattering. In addition, although the same
single-temperature C atmosphere model applied to the CCOs in Cas A and
G353.6--0.7 produces a satisfactory fit to the spectrum of PSR
J1852+0040, the implied emitting area at pulse maximum is comparable
to the total neutron star surface area. This finding is difficult to
reconcile with the observed large amplitude X-ray pulsations,
suggesting that similar results obtained for other CCOs, especially
G353.6--0.7, may not be valid as well.

In future investigations, it is important to employ realistic
atmospheres in modeling the X-ray emission from PSR J1852+0040. As
noted previously, since the exact magnetic field and chemical
composition for CCOs, in general, are quite uncertain, a wide variety
of models need to be considered.  Moreover, substantially deeper X-ray
observations are needed to better constrain the energy-dependence and
reveal the small-scale details of the pulse profile, especially in
the pulse peak and trough, to further constrain the details of the
surface heat distribution and, by extension, the magnetic field
topology.  In the theoretical realm, it is crucial to investigate the
surface heat signatures of non-standard magnetic field configurations
(e.g.,~non-star-centered and non-axisymmetric), since they appear to
be required to reproduce the phenomenology of PSR J1852+0040.

\acknowledgements I thank E.~V.~Gotthelf for helpful tips regarding
the data reduction, J.~P.~Halpern for insightful discussions, and the
anonymous referee whose helpful comments resulted in substantial
improvements in the manuscript.  This project was supported by NASA
Astrophysics Data Analysis Program (ADAP) grant NNX12AE24G.  The work
presented was based on observations obtained with \textit{XMM-Newton},
an ESA science mission with instruments and contributions directly
funded by ESA Member States and NASA.  This research has made use of
the NASA Astrophysics Data System (ADS) and data obtained from the
High Energy Astrophysics Science Archive Research Center (HEASARC),
provided by NASA's Goddard Space Flight Center.

Facilities: \textit{XMM-Newton} (EPIC)


\end{document}